\documentclass[12pt]{article}
\usepackage{graphics}
\begin{document}

\begin{center}
{\large\bf PROPAGATION OF THE BURST OF RADIATION IN EXPANDING AND RECOMBINING 
UNIVERSE: THOMSON SCATTERING}
\end{center}

\begin{center}
S.I.Grachev$^1$, V.K.Dubrovich$^{2,}$\footnote[3]{E-mail: dvk47@mail.ru}
\end{center}

\begin{center}
{\it
$^1$Sobolev Astronomical Institute of Saint-Petersburg State University\\
$^2$Saint-Petersburg department of Special Astrophysical Observatory of Russian
Academy of Sciences}\\
\end{center}

{\bf Abstract.} Within the framework of
a flat cosmological model a propagation of an instantaneous burst of 
nonpolarized isotropic
radiation is considered from the moment of its beginning at some initial
redshift $z_0$ to the moment of its registration now (at $z=0$). Thomson 
(Rayleigh) scattering by free electrons is considered as the only source of 
opacity. Spatial distributions of the mean (over directions) radiation 
intensity are calculated as well as angular distributions of radiation 
intensity and polarization at some different distanses from the center of the
burst. It is shown that for redshifts $z_0$ large enough ($z_0\geq 1400$) the
profile of the mean intensity normalized to the total number of photons 
emitted during the burst weakly depends on initial conditions (say the
moment $z_0$ of the burst, the width and shape of initial radiation 
distribution in space). As regards angular distributions
of intensity and polarization they turn to be rather narrow (3 -- 10 arcmin)
while polarization can reach 70\%. On the average an expected polarization
can be about 15\%.

\bigskip

\noindent {\it Key words:} cosmology, early Universe, cosmological 
recombination, radiative transfer, Thomson scattering.

\begin{center}
INTRODUCTION
\end{center}

Investigations of cosmic microwave background (CMB) both theoretical and
experimental ones continue extensively all over the world. Achivements on this
way are enormous. Information about fundamental parameters of the Universe is
obtained. Contribution of completely new kind of matter inaccesible for
investigation in ground based laboratories is discovered and measured. The
main conclusions of the theory of origin of large scale structure of visible
matter distribution in the Universe are confirmed. However the prospects of 
future investigations seem to be still more grandiose ones. In particular 
completely
new discoveries are possible in the course of thorough studying of small scale
CMB temperature fluctuations. Our paper is devoted to one of the mechanisms of
such fluctuations.

In standard cosmological model fluctuations of matter density are possible
on different scales -- from very large to comparatively small ones. An
interaction of these fluctuations with CMB during Universe expansion and 
cooling determines observed pattern of CMB temperature variations ($\Delta 
T/T$)
at the moment of hydrogen recombination at a redshift $z\approx 1000$.
Scales smaller then someones "damp" i.e. give very small values of $\Delta 
T/T$. This takes place for different reasons, namely through decrease of 
peculiar velocity
of matter density fluctuation (cloud) due to radiation viscosity which leads to
decrease of a Doppler shift of photons frequency; through decrease of cloud
size relative to horizon and decrease of contribtution of Sacks--Wolfe effect;
through decrease of cloud optical thickness for Thomson scattering due to 
electron
density decrease. However decrease of amplitude of the own matter density
fluctuation  takes place in this case for sonic waves only! At the same time
for some class of fluctuations their contrast relative to surrounding matter
increases. All these results are true for standard small density fluctuations.
But we consider an evolution of $\Delta T/T$ for nonstandard models of origin
of energy release on small scales. In particular this can be primary black 
holes, domains of different unstable forms of matter etc. at the epoch before
hydrogen recombination. In such an aspect this problem was formulated in the
paper by Dubrovich (2003). In the same paper an evolution of spatial 
distribution of "superequilibrium" radiation intensity was described 
qualitatively. Conclusion was obtained that an angular size of $\Delta T/T$
fluctuation observed now did not depend on a redshift of radiation source.

The aim of the present paper is a correct calculation of this process within 
the
framework of standard theory of radiative transfer. As a model we use a source
of sufficiently small size radiating during some small interval of time
(instantaneously in particular). We assume that this radiation does not
have any influence on the matter parameters (e.g. electron temperature and
number density) during the whole process of scattering. Within the framework 
of the
present paper we do not discuss an absolute quantity of scattered radiation
intensity as well as its spectrum. Separate paper will be devoted to this
question. We stress once more that we consider here completely different
type of temperature inhomogeneity as compared with the standard scenario.
Our source of inhomogeneity is not a small fluctuation of matter density
described in the course of its evolution by equations of hydrodynamics and
thermodynamics. Therefore conclusions of the standard theory of small scale
fluctuations damping (Silk damping) are not applicable to it.

\begin{center}
BASIC EQUATIONS
\end{center}

We consider nonstationary radiation transfer in homogeneous expanding Universe
starting from spherically-symmetrical initial distribution of radiation
intensity. For a flat model of the Universe the corresponding equation of
radiation transfer for photon occupation number $\vec{n}$ has the form 
(Nagirner, Kirusheva, 2005):
\begin{equation}
\frac{\partial \vec{n}}{\partial \eta}+\mu\frac{\partial\vec{n}}{\partial \rho}
+\frac{1-\mu^2}{\rho}\frac{\partial \vec{n}}{\partial \mu}-\frac{a\,\nu}{c}H
\frac{\partial \vec{n}}{\partial\nu}=-a\,k\,(\vec{n}-\vec{s}),
\label{eq1}
\end{equation}
where $\eta$ is a conformal time ($d\eta=cdt/a(\eta)$, $c$ is the speed of 
light, 
$t$ is a time, $a=a(\eta)$ is a scale factor), $\rho$ is a spatial coordinate, 
$H=H(\eta)$ is a Hubble factor, $\vec{n}=(c^2/2h\nu^3)(I,Q)^{\rm T}$,
$I=I(\rho,\mu,\eta, \nu)$ and $Q=Q(\rho,\mu,\eta,\nu)$ are Stokes parameters
for radiation at a frequency $\nu$ propagating at an angle $\vartheta=
\arccos\mu$ to radial direction, dimensionless vector source function is 
$\vec{s}=(s_I,s_Q)^{\rm T}=(c^2/2h\nu^3)(S_I,S_Q)^{\rm T}=(c^2/2h\nu^3)
\vec{\varepsilon}/k$ and $\vec{\varepsilon}=\vec{\varepsilon}(\rho,\mu,\eta,
\nu)$ and $k=k(\rho,\mu,\eta,\nu)$ are emission and absorption coefficients 
accordingly. It should be noted that $k=k(\eta)$ for Thomson scattering in a
homogeneous medium. Here and further $^{\rm T}$ means transpose.

Below we assume scale factor $a$ to be equal 1 at the present epoch ($z=0$) so
that $a(z)=1/(1+z)$ where $z$ is a redshift. Therefore spatial coordinate
$\rho$ in eq. (\ref{eq1}) is a distance from a center of symmetry measured in
the present epoch (at $z=0$). At an arbitrary $z$ the corresponding distance 
is $r=\rho\,a(z)=\rho/(1+z)$. Further since Thomson scattering is neutral one
than radiation frequency $\nu$ changes due to cosmological redshift only:
$\nu=\nu_0/a(z)=\nu_0(1+z)$, where $\nu_0$ is a frequency at the present epoch.
Therefore we may use $\nu_0$ instead of $\nu$ as a frequency variable and we
redefine the source function (Nagirner, Kirusheva, 2010) as follows: 
\begin{equation}
\vec{n}(\rho,\mu,\eta,\nu)=\vec{n}_*(\rho,\mu,\eta,\nu_0)\equiv
\vec{n}_(\rho,\mu,\eta).
\label{nusep}
\end{equation}
Then frequency variable disappears and eq. (\ref{eq1}) takes the form
\begin{equation}
\frac{\partial \vec{n}}{\partial \eta}+\mu
\frac{\partial \vec{n}}{\partial \rho}+\frac{1-\mu^2}{\rho}\frac{\partial
\vec{n}}{\partial\mu}=-k(\eta)a(\eta)\left[\vec{n}
(\rho,\mu,\eta)-\vec{s}(\rho,\mu,\eta)\right].
\label{maineq}
\end{equation}
In the case of monochromatic scattering and in the presence of axial symmetry
of radiation field the vector source function in the righthand side of this
equation is written in the form 
\begin{equation}
\vec{s}(\rho,\mu,\eta)=(1/2)\int_{-1}^{+1}d\mu'
\hat{R}(\mu,\mu')\vec{n}(\rho,\mu',\eta),
\label{Sfun}
\end{equation}
where phase matrix $\hat{R}(\mu,\mu')$ describes radiation redistribution 
over directions and polarizations in a single scattering. For Thomson 
(Rayleigh) scattering phase matrix can be factorized as follows (see e.g.
Ivanov, 1995):
\begin{equation}
\hat{R}(\mu,\mu')=\hat{A}(\mu)\hat{A}^{\rm T}(\mu'),
\end{equation}
where 
\begin{equation}
\hat{A}(\mu)=\left(\begin{array}{cc}  1 & \frac{1}{2\sqrt{2}}(1-3\mu^2)\\
                                 0 &  \frac{3}{2\sqrt{2}}(1-\mu^2)\\
				 \end{array}\right).
\end{equation}
In this case eq. (\ref{Sfun}) takes the form
\begin{equation}
\vec{s}(\rho,\mu,\eta)=\hat{A}(\mu)\vec{s}_*(\rho,\eta),
\label{SA}
\end{equation}
where
\begin{equation}
\vec{s}_*(\rho,\eta)=(1/2)\int_{-1}^{+1}\hat{A}^{\rm T}(\mu')\vec{n}(\rho,
\mu',\eta)d\mu'.
\label{Sast}
\end{equation}

Eq. (\ref{maineq}) looks like an ordinary nonstationary equation of
monochromatic radiation transfer in a medium with the volume absorption 
coefficient
\begin{equation}
\alpha(\eta)=k(\eta)a(\eta),
\label{alpha}
\end{equation}
which depends on time only. Therefore $\eta/c$ can be considered as an 
ordinary time and $\rho$ as an ordinary radial distance in the problems of
nonstationary radiative transfer in spherically-symmetrical extended media.
Thus we can use the method suggested earlier by Ivanov (1995) for solving 
problems of stationary transfer in plane-parallel media with generalized
Rayleigh scattering. The method consists in reducing the problem to getting
and solving an integral equation for so called reduced source function defined
by eq. (\ref{Sast}) which does not depend on angle variable. After that 
radiation field with its angular structure is defined from the formal (i.e. 
with the known source function) solution of initial eq. (\ref{maineq}).

To obtain above mentioned formal solution we consider propagation of
radiation along a ray intersecting radial direction at an angle $\vartheta=
\arccos\mu$ at a distance $\rho$ from the center of symmetry. Let us introduce
coordinate $l$ measured along a ray from the point nearest to the center of
symmetry in direction of radiation propagation.
The lefthand side of eq. (\ref{maineq}) is nothing else than derivative along 
the ray direction so that this equation is rewritten as
\begin{equation}
d\vec{n}/dl=-\alpha(l)[\vec{n}(l)-\vec{s}(l)].
\label{meql}
\end{equation}
Integrating this equation we obtain
\begin{equation}
\vec{n}(l)=\vec{n}_0(l_0)e^{-\int_{l_0}^l\alpha(l')dl'}+\int_{l_0}^l
\alpha(l')\vec{s}(l')e^{-\int_{l'}^l\alpha(l'')dl''}dl'.
\label{fsoll}
\end{equation}
Here $l_0$ is a coordinate of a point farthest from the observation point $l$
but yet capable of making a contribution to radiation in this point at a given 
time $\eta$. Each point $l'$ on a ray is defined by radial distance $\rho'$ and
by an angle $\arccos\mu'$ at which the ray intersects radial direction.
It follows from geometry of the problem that
\begin{equation}
l=\rho\mu,\quad l'=\rho'\mu',\quad l_0=\rho_0\mu_0,\quad \rho\sqrt{1-\mu^2}=
\rho'\sqrt{1-\mu'^2}
\label{rrll}
\end{equation}
and
\begin{equation}
l-l_0=\eta,\quad l-l'=\eta-\eta'.
\label{ll}
\end{equation}
According to the last of these equations one can turn to integration over time
in eq. (\ref{fsoll}) because $dl'=d\eta'$. As a result the formal solution 
(\ref{fsoll}) can be written as follows
\begin{equation}
\vec{n}(\rho,\mu,\eta)=\vec{n}_0(\rho_0,\mu_0)e^{-\int_0^\eta\alpha(\eta'')
d\eta''}+\int_0^\eta\alpha(\eta')\hat{A}(\mu')\vec{s}_*(\rho',
\eta')e^{-\int_{\eta'}^\eta\alpha(\eta'')d\eta''}d\eta',
\label{fsol}
\end{equation}
taking into account eqs. (\ref{SA}), (\ref{rrll}) and (\ref{ll}). Here vector
$\vec{n}_0(\rho,\mu)$ defines initial (at the moment $\eta=0$) distribution
over distances and angles and
\begin{equation}
\rho_0=\sqrt{\rho^2-2\rho\mu\eta+\eta^2},\quad \rho_0\mu_0=\rho\mu-\eta,
\label{rhomu0}
\end{equation}
\begin{equation}
\rho'=\sqrt{\rho^2-2\rho\mu(\eta-\eta')+(\eta-\eta')^2},\quad \rho'\mu'=
\rho\mu-\eta+\eta'.
\label{rhomup}
\end{equation}

It follows from the formal solution (\ref{fsol}) that one can introduce 
dimensionless time
\begin{equation}
u=\int_0^\eta\alpha(\eta')d\eta'=c\sigma_e\int_0^t n_e(t')dt'=
c\sigma_e\int_z^{z_0} \frac{n_e(z')}{(1+z')H(z')}dz,'
\label{udef}
\end{equation}
which has a sense of an optical distance (by Thomson scattering) between 
moments $z$ and $z_0$. Here redshift $z_0$ corresponds to the initial moment
of time: $u=\eta=t=0$ for $z=z_0$, $n_e$ is an electron number density, 
$\sigma_e=6.65\cdot 10^{-25}$ см$^2$ is the cross-section of Thomson
scattering. For conformal time $\eta$ we have equation
\begin{equation}
\eta=c\int_0^{t}dt'/a(t')=c\int_z^{z_0}dz'/H(z'),
\label{etaz}
\end{equation}
which can be used to relate $u$ with $\eta$ by calculation both of them on
the same grid of redshifts $z$.
With the new time variable $u$ the formal solution (\ref{fsol}) takes the 
form
\begin{equation}
\vec{n}(\rho,\mu,u)=\vec{n}_0(\rho_0,\mu_0)e^{-u}+\int_0^u\hat{A}(\mu')
\vec{s}_*(\rho',u')e^{u'-u}du'.
\label{fsolu}
\end{equation} 
Substitution of eq. (\ref{fsolu}) into the righthand side of eq. (\ref{Sast}) 
gives the following equation for $\vec{s}_*(\rho,u)$:
\begin{equation}
\vec{s}_*(\rho,u)=\vec{s}_0(\rho,u)+\frac{1}{2\rho}
\int_0^u e^{u'-u}
\frac{du'}{\eta-\eta'}\int_{|\rho-\eta+\eta'|}^{\rho+\eta-\eta'}
\hat{A}^{\rm T}(\mu)\hat{A}(\mu')\vec{s}_*(\rho',u')\rho'd\rho',
\label{intequ}
\end{equation}
where arguments of matrices $\hat{A}^{\rm T}$ and $\hat{A}$ are equal
(according to eq. (\ref{rhomup})) to
\begin{equation}
\mu=[\rho^2-\rho'^2+(\eta-\eta')^2]/[2\rho(\eta-\eta')],\quad
\mu'=[\rho^2-\rho'^2-(\eta-\eta')^2]/[2\rho'(\eta-\eta')].
\label{mumup}
\end{equation}
Further the primary source vector in eq. (\ref{intequ}) is
\begin{equation}
\vec{s}_0(\rho,u)=\frac{e^{-u}}{2\rho\eta}
\int_{|\rho-\eta|}^{\rho+\eta}\hat{A}^{\rm T}(\mu)\vec{n}_0(\rho_0,\mu_0)
\rho_0d\rho_0,
\label{s0intu}
\end{equation}
where
\begin{equation}
\mu=(\rho^2-\rho_0^2+\eta^2)/2\rho\eta,\quad
\mu_0=(\rho^2-\rho_0^2-\eta^2)/2\rho_0\eta.
\label{mumu0}
\end{equation}
according to eq. (\ref{rhomu0}).

To deduce the main integral equation (\ref{intequ}) we passed from
integration over $\mu$ to integration over $\rho'$ in the integral term
and to $\rho_0$ in the free term using in the first case the first of
eqs. ({\ref{mumup}) which gives
\begin{equation}
d\mu=-\rho'd\rho'/\rho(\eta-\eta'),
\end{equation}
and in the second case we use the first of eqs. (\ref{mumu0}) which gives
\begin{equation}
d\mu=-\rho_0d\rho_0/\rho\eta.
\end{equation}

In scalar case assuming isotropic scattering the main integral equation 
(\ref{intequ}) has a more simple form
\begin{equation}
s(\rho,u)=s_0(\rho,u)+\frac{1}{2\rho}
\int_0^u e^{u'-u}\frac{du'}{\eta(u)-\eta(u')}\int_{|\rho-\eta(u)+
\eta(u')|}^{\rho+\eta(u)-\eta(u')}s(\rho',u')\rho'd\rho',
\label{inteqsc}
\end{equation}
where
\begin{equation}
s_0(\rho,u)=\frac{e^{-u}}{2\rho\eta}\int_{|\rho-\eta|}^{\rho+
\eta}n_0(\rho_0)\rho_0d\rho_0,
\label{s0isc}
\end{equation}
and from eq. (\ref{fsolu}) we obtain for the formal solution:
\begin{equation}
n(\rho,\mu,u)=n_0(\rho_0)e^{-u}+\int_0^u s(\rho',u')e^{u'-u}du',
\label{fsolsc}
\end{equation}
where
\begin{equation}
\rho_0=\sqrt{\rho^2-2\rho\mu\eta+\eta^2},\quad 
\rho'=\sqrt{\rho^2-2\rho\mu(\eta-\eta')+(\eta-\eta')^2}.
\label{tautau}
\end{equation}
It should be stressed that in this case sourse function coinsides with the
mean (over angle variable) radiation intensity:
\begin{equation}
s(\rho,u)=j(\rho,u)\equiv (1/2)\int_{-1}^1 n(\rho,\mu,u)d\mu
\end{equation}
Here and below the term "intensity" means dimensionless intensity i.e.
photon occupation number.

As an initial condition we assume that at the moment $t=0$ ($\eta=0$)
corresponding to some redshift $z_0$ radiation is nonpolarized and isotropic 
and has spherically-symmetrical distribution:
\begin{equation}
\vec{n}(\rho,\mu,0)=(n_0(\rho),0)^{\rm T},
\end{equation}
where $n_0(\rho)$ is a given function which we take in the form
\begin{equation}
n_0(\rho)=\pi^{-3/2}\rho_*^{-3}\exp[-(\rho/\rho_*)^2]\rightarrow \delta(\rho)/
(4\pi\,\rho^2)\quad \mbox{for}\quad \rho_*\rightarrow 0,
\label{n0rho}
\end{equation}
where $\rho_*$ is a parameter wich can be taken sufficiently small to model 
point source. Obviously $n_0(\rho)$ satisfys to normalization
\begin{equation}
4\pi\int_0^\infty n_0(\rho)\rho^2d\rho=1.
\end{equation}

We solve the problem of propagation of instantaneous burst of radiation in a 
scattering expanding and recombining Universe. Since albedo of Thomson
scattering is equal to 1 the full number of photons emitted in the burst
must be conserved. So that an equality
\begin{equation}
4\pi\int_0^\infty j(\rho,u)\rho^2d\rho=4\pi\int_0^\infty n_0(\rho)\rho^2
d\rho=1.
\label{consrv}
\end{equation}
must be fulfilled at any moment of time. One can make certain that this 
equality follows indeed both from vector equation (\ref{intequ}) and scalar
eqution (\ref{inteqsc}).

\begin{center}
METHOD OF SOLUTION AND MAIN RESULTS
\end{center}

We obtained numerical solutions of vector equation (\ref{intequ}) and
scalar equation (\ref{inteqsc}) by means of their discretization on given
grids over dimensionless time $u$ and over distance $\rho$ measured in Mpc.
It should be noted that the scale of distances $\rho$ corresponds to the
present epoch according to our normalization of scale factor: $a=1$ at $z=0$.
As a main time grid we use a uniform grid over redshift $z$ with the step
$\Delta z=10$ and grids over $u$ and $\eta$ are calculated then using eqs.
(\ref{udef}) and (\ref{etaz}). The function $\vec{s}_*(\rho',u')$ in the
righthand side of integral equation (\ref{intequ}) was approximated as a 
function of $\rho'$ for fixed $u'$ by cubic spline on a uniform grid with the
step $\Delta \rho=1$ Mpc and integral over $\rho'$ was calculated analytically.
Next the whole of integrand (except for exponential factor) in integral over
$u'$ was approximated (as a function of $u'$) by quadratic spline and integral
over $u'$ was calculated analytically as well. At the last time point (for 
$u'=u$, $\eta'=\eta$) the whole integrand in integral over $u'$ including 
the multiplyer $1/2\rho$ turns to be equal to $\hat{B}\vec{s}_*(\rho,u)$ where
$\hat{B}=$diag(1,7/10) is a diagonal matrix. Carrying this term of quadrature
sum from the righthand side of equation to the lefthand one we get in the end
recurrence relation which allows to express the current solution through 
solutions at the preceeding moments of time. To control the process of solution
we check up conservation of the full number of photons. It was fulfilled with
a relative error no more than $10^{-7}$ for scalar equation and no more than
$10^{-5}$ for vector equation.

After the source function was found then angular distributions of radiation
intensity and polarization were calculated numerically from the formal
solution for different distances from the center of the burst and at
different moments ($z$) including the present one ($z=0$). Calculations were
fulfilled for several values of initial moments of time in the range of 
redshifts $z_0$ from 1100 up to 3000.

The width of initial intensity distribution as a function of $\rho$ (see eq.
(\ref{n0rho})) was taken to be $\rho_*=1.5$ Mpc in the scale of distances at 
$z=0$. But in the scale of distances corresponding to the moment of the
burst (at some $z=z_0$) the width of initial distribution becomes (for 
$z_0\gg 1$) much smaller: $r_*=a(z_0)\rho_*=\rho_*/(1+z_0)$.

As regards to another parameters appearing in the problem they enter the Hubble
factor in particular: 
\begin{equation}
H(z)=H_0\sqrt{\Omega_{\Lambda}+(1-\Omega)(1+z)^2+\Omega_{M}(1+z)^3+\Omega_{rel}
(1+z)^4},
\end{equation}
where $H_0=2.4306\cdot 10^{-18}h_0$ s$^{-1}$, $h_0$ is the Hubble constant 
in the units of 75 km/(s$\cdot$Mpc); $\Omega_{M}$, $\Omega_{\Lambda}$ and 
$\Omega_{rel}$ are ratios of densities of matter, dark energy and relativistic
particles (radiation, massless neutrino) to the crytical density $\rho_c=
3H_0^2/(8\pi G)$ at the present epoch; $\Omega=\Omega_{M}+\Omega_{\Lambda}+
\Omega_{rel}$, $\Omega_{rel}=\rho^0_R(1+f_n)/\rho_c$, $\rho^0_R=a_RT_0^4/c^2$
is a mass density of radiation at the present epoch ($T_0$ is the mean 
temperature of CMBR), $f_n$ is a contribution of relativistic (massless) 
neutrino (usually $f_n=0.68$). For the flat model of the Universe we have
$\Omega=1$ and then $\Omega_{M}=1-\Omega_{\Lambda}-\Omega_{rel}$.

Moreover number density of electrons enter the equations. It is measured 
usually
in the units of the total number density $n_{\rm H}$ of hydrogen atoms and
ions: $n_e(z)=x_e(z)n_{\rm H}(z)$, where $x_e(z)$ is so called recombination
history of the Universe and 
\begin{equation}
n_{\rm H}(z)=n_{\rm H}^0(1+z)^3,\quad  n_{\rm H}^0=0.63144\cdot10^{-5}X
\Omega_{\rm B}h_0^2 \,\mbox{см}^{-3},
\end{equation}
where $\Omega_{B}$ is a ratio of barion density to critical density at the
present epoch, $X$ is a hydrogen abundance (by mass). Recombination history 
is calculated separately and enter as an input file. We calculated it using
the programme "recfast.for" (Seager et al., 1999).

We adopted the following values of parameters: $\Omega=1$,
$\Omega_\Lambda=0.7$, $\Omega_{\rm B}=0.04$, $T_0=2.728$ K, hydrogen abundance
$X=0.76$, $\Omega_{rel}=0.85\cdot 10^{-4}$, Hubble constant $H_0=70$ 
km/(s$\cdot$Mpc). 

Fig. 1 shows initial isotropic distribution of radiation. Results of
calculations are shown in Figs. 2 -- 7. Figs. 2 -- 5 are obtained in scalar
approximation and Figs. 6 and 7 -- using exact description of scattering.
Fig. 2 displays propagation of radiation wave generated by the burst at the
epoch ($z=$ 10 and 20) when the Universe was practically transparent for
radiation (we do not consider here reionization connected with the birth of
primary stars). Then radiation propagates freely and an interval between peaks 
in Fig. 2 is equal to the difference of conformal times of the burst for
corresponding redshifts.

In Fig. 3 there are shown distributions of the mean intensity (i.e. source
function) at the present epoch but for different moments ($z_0$) of the burst. 
Certainly all profiles are in the range of distances which do not exceed 
(strictly do not exceed if we neglect the width of initial spatial profile of
the burst) conformal time of the burst (see Tabl. 1) since according to its
definition (eq. (\ref{etaz})) conformal time is equal to a distance passed by 
freely flying photon from the moment of its emission (at some $z_0$) to the 
moment of its registration (at $z=0$) in a fixed scale of distances coinsiding
by our choice with the scale in the present epoch. Also it is clear that
distribution of diffuse radiation should be wider and its maximum should be 
nearer to the center of the burst as compared with straightly passed 
unscattered radiation which should have its maximum at a distance equal to 
conformal time of the burst. While maximum of diffuse radiation is situated
according to our calculations at a distance $\rho=13620$ Mpc irrespective
of the moment $z_0$ of burst beginning. This distance coincides as it is 
expected 
with the conformal time for the last scattering surface at $z\approx 1090$. 
This is illustrated by two-component profile in Fig. 3 for $z_0=1200$ 
containing both diffuse and straghtly passed radiation.

Furthemore it is seen in Fig. 3 that for sufficiently large $z_0\geq 2000$ the
profile of distribution does not depend practically on the time of the burst
beginning. Moreover our calculations show that it does not depend (providing
that the full number of photons conserves) on a characteristic size of the
burst (in our case this is parameter $\rho_*$ in the initial distribution 
(\ref{n0rho})) provided it is not too large (see below). So for sufficiently 
large $z_0$ the profiles of intensity distribution should be in a sence
universal ones i.e. nondepending on initial conditions of the burst as it was 
indicated earlier by Dubrovich (2003). This is connected with the large optical
thickness of the Universe by Thomson scattering at large redshifts (see Tabl. 
1). So that photons emitted during the burst are trapped and a size of  
radiating region changes due to diffusion not too much till the moment of
sufficient clearing of the Universe at $z\approx 1200$ owing to hydrogen
recombination.

It should be noted that in accordance with our solution normalization on the
full number of photons emitted during the burst into space the calculated
quantities have a dimension of inverse volume, namely 1 Mpc$^{-3}$. Therefore
if at the moment of the burst its central intensity (at $\rho=0$) is equal to 
$n_0$ then our profiles must be multiplied by $\pi^{3/2}n_0(\rho_*/1
\mbox{Mpc})^3$ according to the form of our initial profile (eq. (\ref{n0rho})).
But in general case taking into account weak dependence of solution on the
initial conditions one can take as a transitional multiplier e.g.
$(4\pi/3)\overline{n}(\rho_0/1\mbox{Mpc})^3$ where $\rho_0$ is a typical size
of the burst and $\overline{n}$ is intensity averaged in a sphere of this size.

In Fig. 3 one can see also that as redshift $z_0$ approachs to the beginning of
hydrogen recombination the changes of the profile become more and more
appreciable and during recombination when clearing of the medium becomes 
noticable the straightly passed unscattered radiation of the burst appears: at
$z_0=1200$ it is seen as a narrow peak to the right of diffuse maximum and at
$z_0=1100$ it is already dominates (see Fig. 4).

In Fig. 5 one can see angular distributions of radiation intensity at different
distances from the center of the burst and at different moments ($z_0$) 
of the burst (see Tabl. 2). Characteristic feature of these distributions is
their small width (3 -- 10 arcminutes) and the width is smaller on the
leading front than on the rear one and it decreases with the growth of $z_0$.
It should be stressed that owing to such a strong anisotropy the radiation
intensity towards the center of the burst can be almost $10^7$ times larger
than angle averaged intensity (see Tabl. 2 and Fig. 3). It is clear that
degree of diffuse radiation anisotropy is defined by an angle $2\vartheta_d$ 
at which radiating region is seen at the present epoch ($z=0$) on the last 
scattering surface ($z_{\rm ls}\approx 1090$, $\eta_{\rm ls}\approx 13620$ 
Mpc). If $r_d$ is the region radius at that moment then obviously $\mu_d=\cos
\vartheta_d\sim r_d/\eta_{\rm ls}$ for $r_d=\rho_d/(1+z_{ls})\ll\eta_{\rm ls}$.
For straghtly passed radiation the cosine of angle at which the burst (arised
at $z=z_0$) is seen at a distance $\rho$ from the center amounts $\mu_r=\cos
\vartheta_{\rm r}\sim r_*/\rho$ for $r_*=\rho_*/(1+z_0)\ll\rho$ where $\rho_*$
is the initial radius of the burst in the present scale of distances. For
example for the burst at $z_0=1100$ we have $\rho=\eta\approx 1.36\cdot 10^4$ 
Mpc taking $\rho_*=1.5$ Mpc for the initial radius of the burst. So that we 
have $\mu_r\approx 10^{-7}$ and $\vartheta_r\approx 1.5$ arcminutes. In this
case as was already pointed above the main part of radiations comes to us
without scattering on the way. As for diffuse radiation which dominates for
$z_0>1200$ the width of its angular distribution turns out to be larger. Thus 
for $z_0\geq 1600$ the semiwidth of
angular distribution of radiating region at $z=0$ is equal to 3 -- 10 
arcminutes according to our calculations so that its radius at $z=1100$ must
be 2 -- 7 times larger than for the burst at $z_0=1100$ i.e. $\rho_d\approx 
3 - 10$ Mpc in the present scale of distances and $r_d=\rho_d/(1+z_0)$ in the
scale of distances at $z=z_0$. Therefore the characteristic initial radius
of the burst at $z_0\geq 1600$ must be in any case smaller than this estimate
lest it should influence vitally on the propeties of the burst radiation at
the present epoch. 

Further, decrease of anisotropy when passing from the "base" of leading wave 
front over maximum to the "base" of rear front (see Fig. 3 and Tabl. 2) is
explained as follows: photons observed at the largest distances come from the 
nearest part of radiating region in a small solid angle and an effective size
of region giving a contribution to observed radiation grows with the distance
decrease.

\begin{table}
\centering

\begin{tabular}{|c|r|c|}

\hline

$z_0$  & $u$ & $\eta$, Mpc \\
\hline
3000 & 177.23 & 13770 \\
2000 &  67.63 & 13720 \\
1600 &  29.27 & 13686 \\
1400 &  12.74 & 13664 \\
1200 &   3.01 & 13635 \\
1100 &   1.07 & 13618 \\

\hline
\end{tabular}
\caption{Optical distances $u$ and conformal times $\eta$ of the burst from
its beginning at $z_0$ to the present epoch $z=0$.}
\end{table}

\begin{table}
\centering

\begin{tabular}{|c|c|c|c|c|c|}

\hline

N  & $r$, Mpc & $z_0=1600$ & $z_0=2000$ & $z_0=3000$ & $z_0=2000$\\
\hline
1 & 13570 & $1.02\cdot 10^{-6}$ & $1.02\cdot 10^{-6}$ & $1.02\cdot 10^{-6}$ & $1.27\cdot 10^{-6}$\\
2 & 13600 & $6.91\cdot 10^{-6}$ & $6.56\cdot 10^{-6}$ & $6.28\cdot 10^{-6}$ & $7.39\cdot 10^{-6}$\\
3 & 13630 & $1.92\cdot 10^{-5}$ & $1.57\cdot 10^{-5}$ & $1.38\cdot 10^{-5}$ & $1.69\cdot 10^{-5}$\\
4 & 13660 & $6.97\cdot 10^{-7}$ & $7.89\cdot 10^{-7}$ & $8.60\cdot 10^{-7}$ & $9.00\cdot 10^{-7}$\\

\hline
\end{tabular}
\caption{Radiation intensity $n(0)$ towards the center of the burst as 
a function of distance $r$ from the burst center for different moments $z_0$ of
the burst beginning. Last column corresponds to exact description of Thomson
(Raylegh) scattering and preceeding three columns correspond to scalar
approximation.}
\end{table}

Fig. 6 shows change of angular distributions when passing from approximate 
scalar description of Thomson scattering to an exact description taking into
account scattering anisotropy and polarization. Finally Fig. 7 shows that
polarization of radiation can be rather large (up to 70\%) but on angular
distances where radiation intensity becomes already much smaller than towards 
the center of the burst (see preceeding Figure). It should be noted also 
anticorrelation between polarization and anisotropy of radiation. Namely,
when passing from the farthest (from the burst center) point of the wave
profile to the nearest one an anisotropy decreases but polarization grows
(cf. Figs. 6 and 7). The thing is that radiation from farthest points comes
from the nearest to observer small part of emitting region and mainly it 
consists of photons undergone their last scattering almost directly forward 
which changes polarization only slightly. But for the less distant (from the 
burst center)
points an essential contribution comes from photons scattered under 
sufficiently large angles which leads to polarization growth.

\newpage

\begin{center}
CONCLUSIONS
\end{center}

A source of very small size radiating energy in prerecombination epoch 
will be seen now as a some spot on the background of cosmic microwave
radiation. Our calculations confirm initial conclusion made by Dubrovich 
(2003) about nondependence of angular size of this spot on the moment of
the source burst on condition that it is situated at a distance of optimal
visibility. For estimates of radiation intensity it is important that duration
of the source burst also has a very little influence on the size of the spot.
More exactly, it takes place for the time interval before the moment of
the Universe clearing due to hydrogen recombination. Assuming that the burst
radiation has no effect on the medium parameters the burst size does not
depend also from the burst power. Calculated exact profiles of intesity
distribution give an opportunity to determine relation between angular size 
of a spot and physical distance to a source. Very important feature is also 
a presence of radiation polarization in the spot. Polarization is standard
one for the case of star radiation scattered in a shell of gas with free
electrons. The plane of polarization contains the ray of sight and direction
towards the star center.
Degree of polarization grows when moving off a spot center. However the most 
probable observed polarization degree will be about 15\% on average because of
fast brightness decrease to the edge of a spot.   

This research is supported by Russian Foundation for Basic Research
(grant 08-02-000493).

\begin{center}
REFERENCES
\end{center}

\begin{enumerate}
\item Dubrovich V.K., Astronomy Letters {\bf 29}, 6 (2003).
\item Ivanov V.V., Astron. Astrophys. {\bf 303}, 609 (1995).
\item Nagirner D.I., Kirusheva S.L., Astronomy Reports {\bf 49}, 167 (2005).
\item Nagirner D.I., Kirusheva S.L., Astronomy Reports {\bf 54}, 55 (2010).
\item Seager S., Sasselov D.D., Astrophys. J. {\bf 523}, L1 (1999). 
\end{enumerate}

\newpage

\begin{figure}[p]
\vspace*{-4cm}
\centering

\resizebox{1.0\textwidth}{!}{\includegraphics{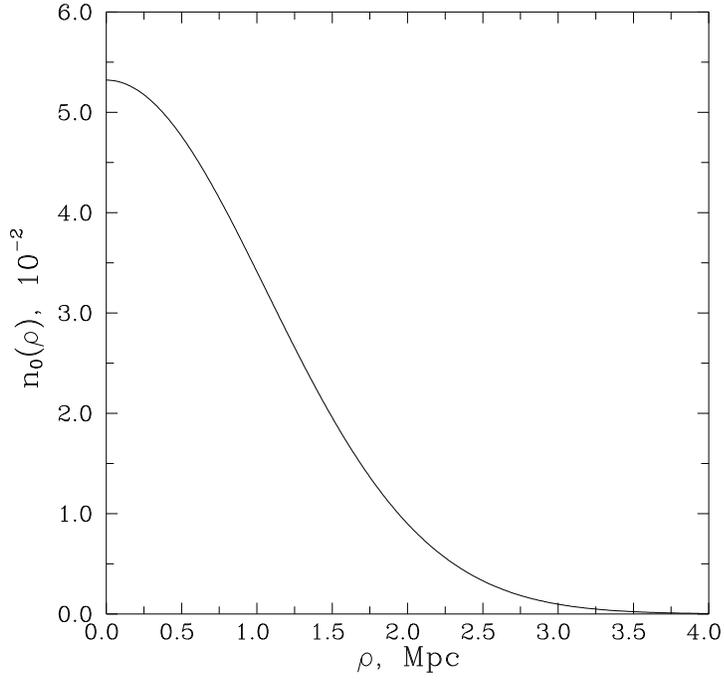}}
\vspace{-11cm}
\caption{Initial (isotropic) distribution of radiation intensity.}

\end{figure}

\begin{figure}[p]

\vspace*{-5cm}
\centering

\resizebox{1.0\textwidth}{!}{\includegraphics{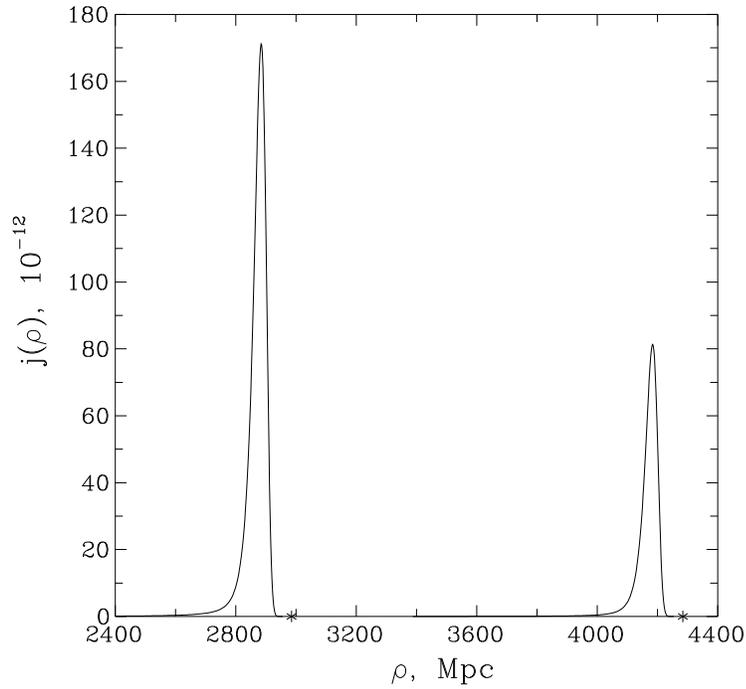}}
\vspace{-11cm}

\caption{Distributions of the mean intensity at $z=10$ (left profile)
and $z=20$ (right profile) ($z_0=2000$). Asteriscs on the abscissa mark 
distances equal to conformal times of the burst for corresponding
$z$.}
\end{figure}

\begin{figure}[p]

\vspace*{-6cm}
\centering

\resizebox{1.0\textwidth}{!}{\includegraphics{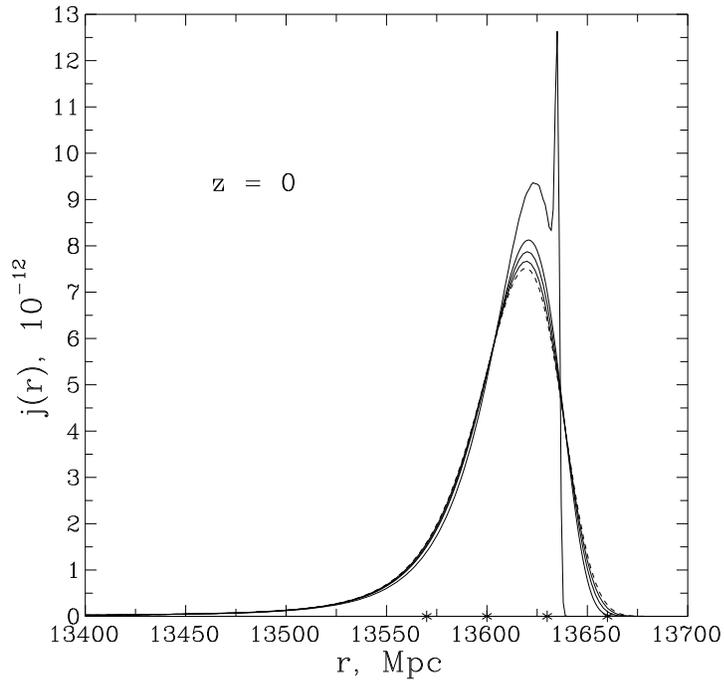}}
\vspace{-11cm}

\caption{Distributions of the mean intensity at the present epoch for different
$z_0$. In the order of maximum intensity growth they correspond to 
$z_0=$3000, 2000, 1600, 1400 and 1200. In the same order the steepness of the 
leading (right) front of distribution grows. Asteriscs on the abscissa mark 
distances indicated in Tabl. 2.}
\end{figure}

\begin{figure}[p]

\vspace*{-6cm}
\centering
\resizebox{1.0\textwidth}{!}{\includegraphics{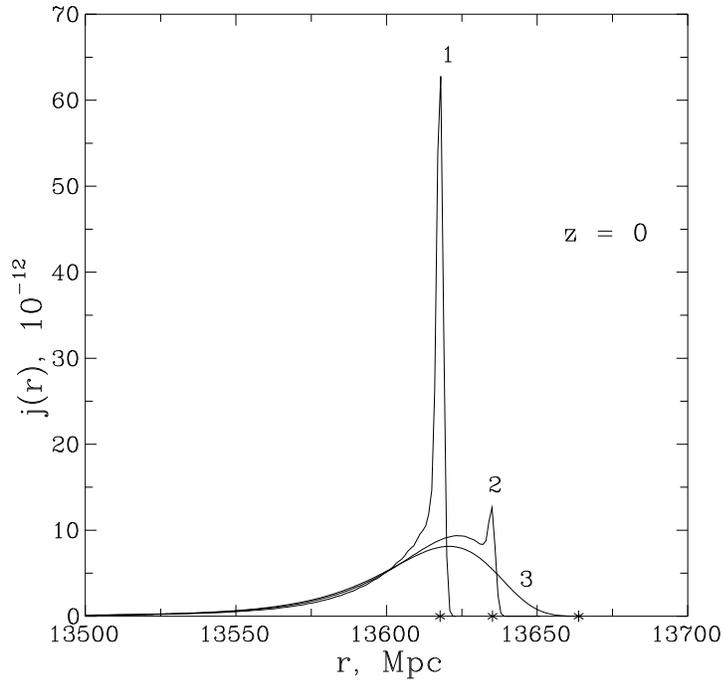}}
\vspace{-11cm}

\caption{The same as in the preceeding Figure but for $z_0=$1100 (1),
1200 (2) and 1400 (3). Asteriscs mark corresponding conformal times of the
burst (see Tabl. 1).}
\end{figure}

\begin{figure}[p]

\vspace*{-6.0cm}
\centering

\resizebox{1.0\textwidth}{!}{\includegraphics{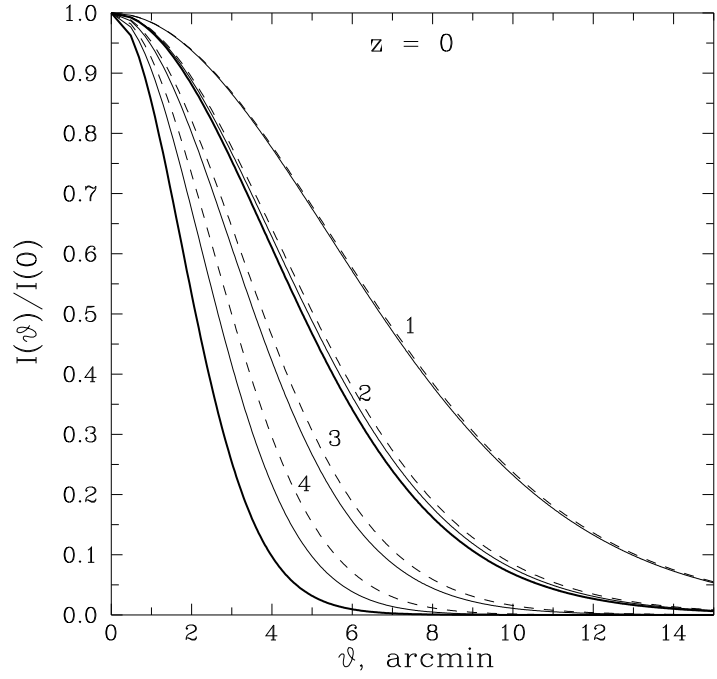}}
\vspace{-11cm}

\caption{Angular distributions of radiation intensity at $z=0$ at different
distances $r$ (see Tabl. 1) from the burst center and for different $z_0$: 
thin continuous lines -- $z_0=2000$, dashed lines -- $z_0=3000$ and thick 
continuous lines -- 
$z_0=1600$ (for two distances -- N $=$ 4 and 2 in Tabl. 2). Here $\vartheta$ 
is an angular distance from the burst center.}

\end{figure}

\begin{figure}[p]

\vspace*{-5.0cm}
\centering

\resizebox{1.0\textwidth}{!}{\includegraphics{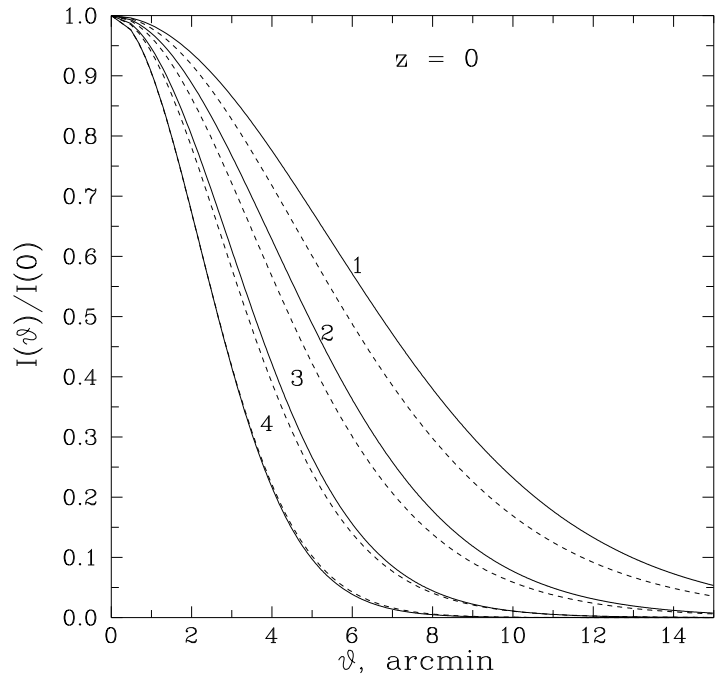}}

\vspace{-11cm}

\caption{The same as in the preceeding Figure for $z_0=2000$ in scalar 
approximation (continuous lines) in comparisson with distributions for exact 
description of Rayleigh scattering (dashed lines).}

\end{figure}

\begin{figure}[p]

\vspace*{-5.0cm}
\centering

\resizebox{1.0\textwidth}{!}{\includegraphics{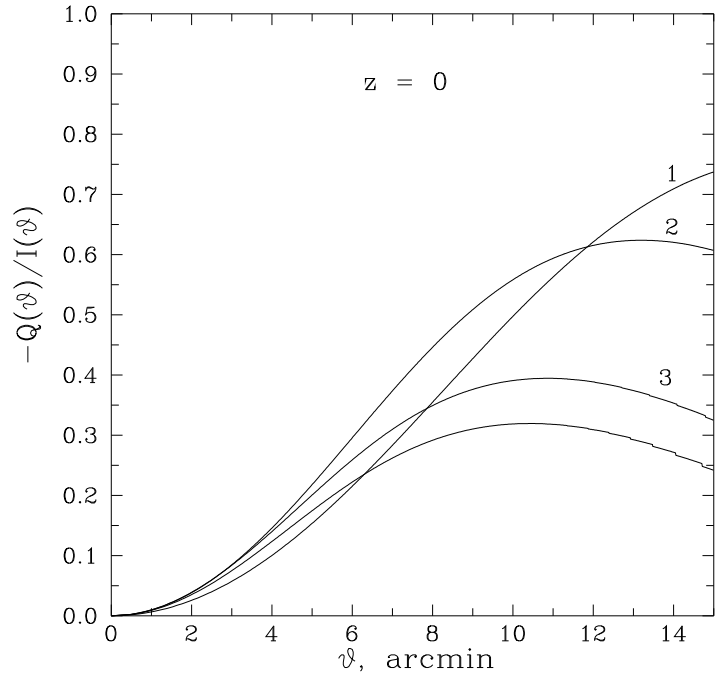}}

\vspace{-11cm}

\caption{Angular distributions of radiation polarization at $z=0$ at different
distances $r$ from the burst center (designations are the same as in the
preceeding Figure). Lowest (with the smallest maximum) curve corresponds to
distance 13640 Mpc.}

\end{figure}

\end{document}